\documentclass[aps,nature,british,superscriptaddress,floatfix,letterpaper,nobalancelastpage,reprint]{revtex4-1}
\usepackage{graphicx} 
\usepackage{color} 
\usepackage{babel}
\usepackage{amsmath} 
\usepackage{hyperref} 
\usepackage{amssymb}
\newcommand{\bra}[1]{\langle{#1}|}

\newcommand{\ket}[1]{\left\lvert #1 \right\rangle}
\DeclareMathOperator{\tr}{Tr}

\newcommand\startsupplement{%
    \makeatletter 
       \setcounter{table}{0}
       \renewcommand{\thetable}{S\arabic\c@table}
       \setcounter{figure}{0}
       \renewcommand{\thefigure}{S\@arabic\c@figure}
    \makeatother}

\usepackage{newfloat}
\DeclareFloatingEnvironment[name={Extended Data Figure}]{suppfigure}

\begin{document}
\title{Implementing a strand of a scalable fault-tolerant quantum computing fabric}

\author{Jerry M. Chow}
\affiliation{IBM T.J. Watson Research Center, Yorktown Heights, NY 10598, USA}
\author{Jay M. Gambetta}
\affiliation{IBM T.J. Watson Research Center, Yorktown Heights, NY 10598, USA}
\author{Easwar Magesan}
\affiliation{IBM T.J. Watson Research Center, Yorktown Heights, NY 10598, USA}
\author{Srikanth J. Srinivasan}
\affiliation{IBM T.J. Watson Research Center, Yorktown Heights, NY 10598, USA}
\author{Andrew W. Cross}
\affiliation{IBM T.J. Watson Research Center, Yorktown Heights, NY 10598, USA}
\author{David W. Abraham}
\affiliation{IBM T.J. Watson Research Center, Yorktown Heights, NY 10598, USA}
\author{Nicholas A. Masluk}
\affiliation{IBM T.J. Watson Research Center, Yorktown Heights, NY 10598, USA}
\author{B. R. Johnson}
\affiliation{Raytheon, BBN Technologies, Cambridge, MA 02138, USA}
\author{Colm A. Ryan}
\affiliation{Raytheon, BBN Technologies, Cambridge, MA 02138, USA}
\author{M. Steffen}
\affiliation{IBM T.J. Watson Research Center, Yorktown Heights, NY 10598, USA}

\date{\today}
\maketitle

\noindent\textbf{\boldmath Quantum error correction (QEC) is an essential step towards realising scalable quantum computers. Theoretically, it is possible to achieve arbitrarily long protection of quantum information from corruption due to decoherence or imperfect controls, so long as the error rate is below a threshold value~\cite{Aharonov1997,Preskill1998}. The two-dimensional surface code (SC) is a fault-tolerant error correction protocol~\cite{Bravyi1998,Raussendorf2007} that has garnered considerable attention for actual physical implementations, due to relatively high error thresholds $\sim$$1\%$, and restriction to planar lattices with nearest-neighbour interactions. Here we show a necessary element for SC error correction: high-fidelity parity detection of two code qubits via measurement of a third syndrome qubit. The experiment is performed on a sub-section of the SC lattice with three superconducting transmon qubits~\cite{koch_charge-insensitive_2007}, in which two independent outer code qubits are joined to a central syndrome qubit via two linking bus resonators~\cite{majer_coupling_2007}. With all-microwave high-fidelity single- and two-qubit nearest-neighbour entangling gates, we demonstrate entanglement distributed across the entire sub-section by generating a three-qubit Greenberger-Horne-Zeilinger (GHZ) state with fidelity $\sim$$94\%$. Then, via high-fidelity measurement of the syndrome qubit, we deterministically entangle the otherwise un-coupled outer code qubits, in either an even or odd parity Bell state, conditioned on the syndrome state. Finally, to fully characterize this parity readout, we develop a new measurement tomography protocol to obtain a fidelity metric ($90\%$ for odd and $91\%$ for even). Our results reveal a straightforward path for expanding superconducting circuits towards larger networks for the SC and eventually a primitive logical qubit implementation.}

The initial work on QEC was to show that a threshold exists which makes fault-tolerant quantum computing possible~\cite{Aharonov1997,Knill1998}. However, most of these proposals were purely theoretical studies which would be impractical to implement in a physical system. 
Knill's C4/C6 architecture~\cite{knill_nature_2005} showed it was possible to have a pseudo-threshold as high as $3\%$ but with significant overhead. 
The SC is a topological error-correcting code \cite{Bravyi1998,Dennis2002}, and the SC protocol \cite{Raussendorf2007} achieves high thresholds with reasonable overheads in
the physically setting of a two-dimensional lattice of qubits supporting nearest-neighbour interactions. SCs have now been proposed for a number of physical qubit systems \cite{DiVincenzo2009,fowler_surface_2012}; however, a complete demonstration remains an outstanding challenges of technical integration and implementation.

\begin{figure}[t!]
\centering
\includegraphics[width=0.47\textwidth]{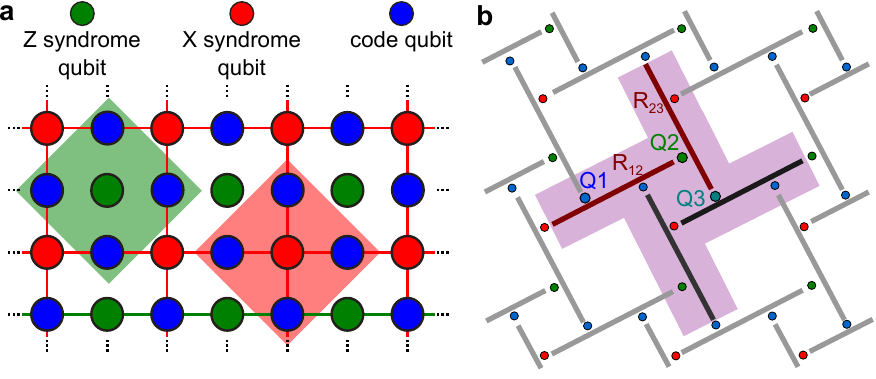}
\caption{\label{fig:1} \textbf{Two-dimensional surface code and skew-symmetric layout.} \textbf{a}, Lattice of nearest-neighbour qubits for realizing the surface code. The lattice contains three flavours of qubits, code (blue circles), $X$-syndrome (red circles), and $Z$-syndrome (green circles) qubits. The lattice is composed of two types of plaquettes, an $X$-plaquette (red-shaded diamond) and a $Z$-plaquette (green-shaded diamond). Critical to the surface code is performing CNOT operations between code qubits and their neighbouring syndrome qubits, followed by measuring the $Z$- and $X$-syndrome qubits to determine the parity of the code qubits. \textbf{b}, The surface code lattice can be mapped into a physical skew-symmetric layout using superconducting qubits (colored circles as in \textbf{a}) coupled to resonators (gray bars). Here each qubit need only connect to 2 bus resonators instead of the required 4 in the strict square-lattice approach. A full eight-qubit plaquette is indicated by the purple-shaded region. The half-plaquette device studied here is labeled, with three qubits (Q1, Q2, Q3), and two bus resonators $\text{R}_{12}$ and $\text{R}_{23}$. Note that Q3 is also a code qubit, but colored in teal for clarity to be distinguished from Q1.}
\end{figure}

\begin{figure*}[htb!]
\centering
\includegraphics{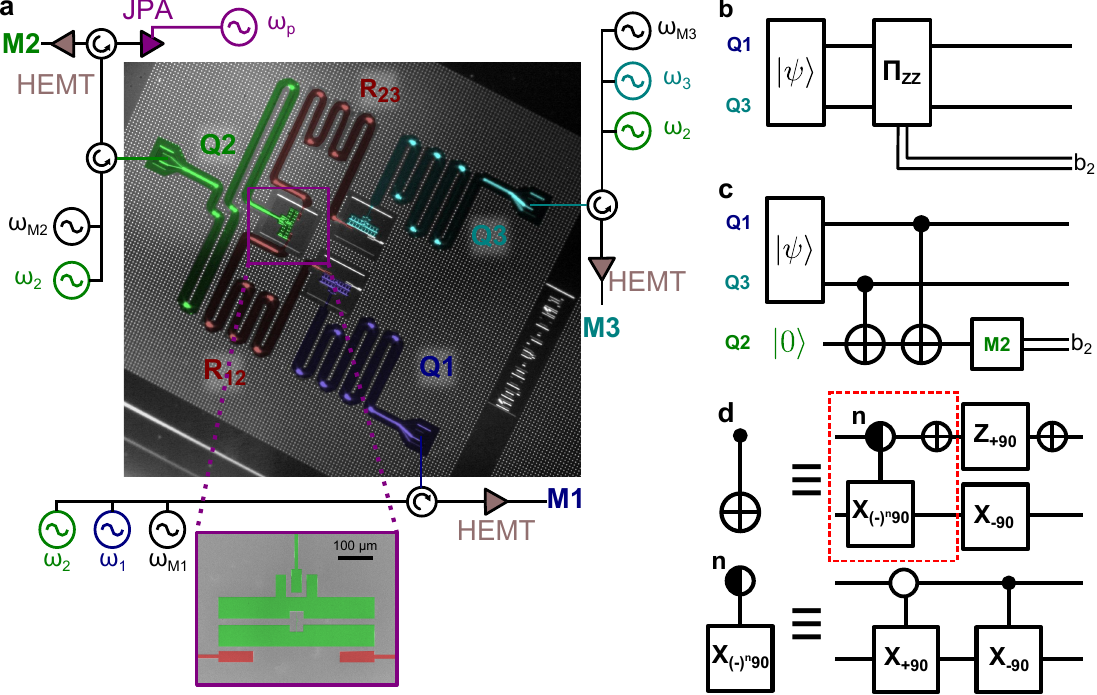}
\caption{\label{fig:2} \textbf{Half-plaquette device schematic and parity check quantum circuits.} \textbf{a}, The  optical image of the half-plaquette device shows in false color all the different components of the device: 3 qubits, Q1 (blue), Q2 (green), and Q3 (teal), each with individual readout resonators, and 2 bus resonators (maroon) $\text{R}_{12}$ and $\text{R}_{23}$. Each transmon qubit (zoom view inset) is independently addressed via its corresponding readout resonator, with single-qubit gates applied on resonance with each qubit at $\omega_i$, $i \in [1,2,3]$ and readout performed at the measurement frequencies $\omega_{\text{M}i}$. Whereas Q1 and Q3 readout signals are only amplified through High-Electron Mobility Transistors (HEMTs), the Q2 readout is reflected off a Josephson parametric amplifier (JPA) stage first before going on to a HEMT. Two-qubit gates are performed in the cross-resonance scheme, applying $\omega_2$ on both control qubits, Q1 and Q3. \textbf{b}, The parity check protocol (PCP) for qubit Q1 and Q3 where the $Z$-parity operator $\Pi_{ZZ}$ is applied, giving a single classical bit of information $b_2$ (double lines indicate classical channel). \textbf{c}, The quantum circuit which implements the $Z$-parity check consists of a pair of CNOT gates from the code qubits (Q1 and Q3) to the syndrome (Q2) followed by a measurement M2 which gives the classical bit $b_2$. \textbf{d}, The CNOT can be decomposed into the $ZX_{90}$ gate and single-qubit rotations. Using the cross-resonance microwave interaction, we have at our disposal the gate combination boxed in dashed red, composed of a $ZX_{90}$ followed by a NOT (or $X$) gate on the control qubit. The $n$ in the depiction of the $ZX_{90}$ gate can be either 0 or 1, indicating the state-dependent rotation.}
\end{figure*}

With on-going improvements to coherence times~\cite{Paik2011, rigetti_superconducting_2012,chang_improved_2013,Barends2013}, gate~\cite{Chow2012PRL,corcoles_process_2013} and readout fidelities~\cite{Bergeal2010,riste_initialization_2012,johnson_heralded_2012} at or approaching threshold values delimited by SCs, superconducting qubits are prime candidates for scaling up towards a fault-tolerant architecture. A scheme for building a network of superconducting qubits employs microwave resonators as the links, as it has been shown that resonators can be used as quantum buses to mediate interactions between qubits~\cite{majer_coupling_2007}, and that multiple resonators can be coupled to a single qubit~\cite{johnson_quantum_2010}. In the future, larger superconducting qubit systems and networks will also leverage circuit integration techniques which come along with a solid-state architecture.

The SC layout is comprised of qubits arranged in a square lattice as shown in Fig.~\ref{fig:1}\textbf{a}. The qubits in the lattice come in two distinct flavours, either code qubits which carry logical information, or syndrome qubits which are used to measure stabilizer operators of surrounding code qubits. In Fig.~\ref{fig:1}\textbf{a}, syndrome qubits can be used as either $X$-parity checks (red circles) or $Z$-parity checks (green circles) of four surrounding code qubits (blue circles). The five-qubit block consisting of $X$- ($Z$-) syndrome and four surrounding code qubits defines a unit cell $X$- ($Z$-) plaquette and is shaded in red (green). Performing a round of error-correction in the SC consists of applying controlled-NOT (CNOT) gates to map the parity of the surrounding code qubits into the state of the syndrome qubits. A subsequent measurement of the syndrome state determines this parity.

Using superconducting resonators as the links of the lattice, it is possible to construct a SC architecture using superconducting qubits in a square lattice, with a qubit at each vertex. This requires the ability to couple a single qubit to four resonators. As it is also important to be able to read out and address the qubits individually, this can result in an additional fifth resonator per qubit. However, another approach to topologically achieve the same SC is to use the skew-symmetric square lattice shown in Fig.~\ref{fig:1}\textbf{b}. This only requires that a single qubit be coupled to two buses, and a third readout resonator. Such an architecture is commensurate with experiments already demonstrated, where a single qubit can be coupled to two separate resonators~\cite{johnson_quantum_2010}, a single bus has been used to couple up to 3 qubits~\cite{DiCarlo2010}, and an independent readout can be used to measure a single qubit that has been entangled with a separate qubit via a bus resonator~\cite{groen_partial-measurement_2013}. In the skew-symmetric lattice, a full plaquette cell consists of 8 qubits and 4 bus resonators, and is indicated in the purple-shaded region of Fig.~\ref{fig:1}\textbf{b}. The experiments presented in this manuscript are performed effectively on a `half-plaquette' sub-section, consisting of 3 qubits (all with individual readout resonators) and 2 bus resonators, where we demonstrate all necessary gate operations and measurements that comprise the SC protocol.

The half-plaquette device (Fig.~\ref{fig:2}\textbf{a}) contains three single-junction transmon qubits connected by two coplanar-waveguide (CPW) resonators serving as the buses, and each qubit is coupled to its own separate CPW resonator for independent readout and control. Here we label the code qubits Q1 and Q3, and the middle qubit Q2 serves as the syndrome. A simplified control and readout diagram is also shown in Fig.~\ref{fig:2}\textbf{a}, with all microwave sources for single- and two-qubit gates and independent readout indicated. The syndrome Q2 is read out with the assistance of a Josephson parametric amplifier~\cite{hatridge_dispersive_2011,johnson_heralded_2012} (JPA) for high-fidelity single-shot state discrimination, which is a critical component for demonstrating the SC parity check protocol. All device parameters and relevant coherence times are given in the Methods Summary and a complete schematic is given in Extended Data Fig.~\ref{fig:S1}. All single-qubit controls are performed in 40 ns, and characterized to $>99.7\%$ average gate fidelity via simultaneous randomized benchmarking~\cite{gambetta_characterization_2012} (Table~S1).

At the crux of the SC protocol is the detection of the parity of the code qubits. Figure~\ref{fig:2}\textbf{b} shows the parity check protocol (PCP) in a circuit depiction on an arbitrary state $\ket{\psi}$ of Q1 and Q3, with the parity being indicated through the classical detection of an indication bit, $b_2$. The PCP is realized in a system of 3 qubits via the quantum circuit shown in Fig.~\ref{fig:2}\textbf{c}, where the parity of the Q1 and Q3 state $\ket{\psi}$ is mapped onto the syndrome Q2 via two CNOT gates, between Q1 and Q2, Q3 and Q2, and then subsequently the classical indication bit $b_2$ is obtained via a quantum measurement of Q2.

High-fidelity CNOT entangling gates are critical for the PCP. To realise these CNOTs in our device, we implement entangling $ZX_{90}$ gates using the cross-resonance interaction~\cite{Chow2011}. By driving Q1 (Q3) at the Q2 transition frequency $\omega_2$, we use a simple decoupling sequence~\cite{corcoles_process_2013} to implement the two-qubit Clifford generator $ZX_{90}$ gate between Q2 and Q1 (Q3). The $ZX_{90}$ gate is equivalent to a CNOT up to single-qubit rotations, as illustrated in Fig.~\ref{fig:2}\textbf{d}, and thus can be used interchangeably in the PCP.

\begin{figure*}[htbp!]
\centering
\includegraphics{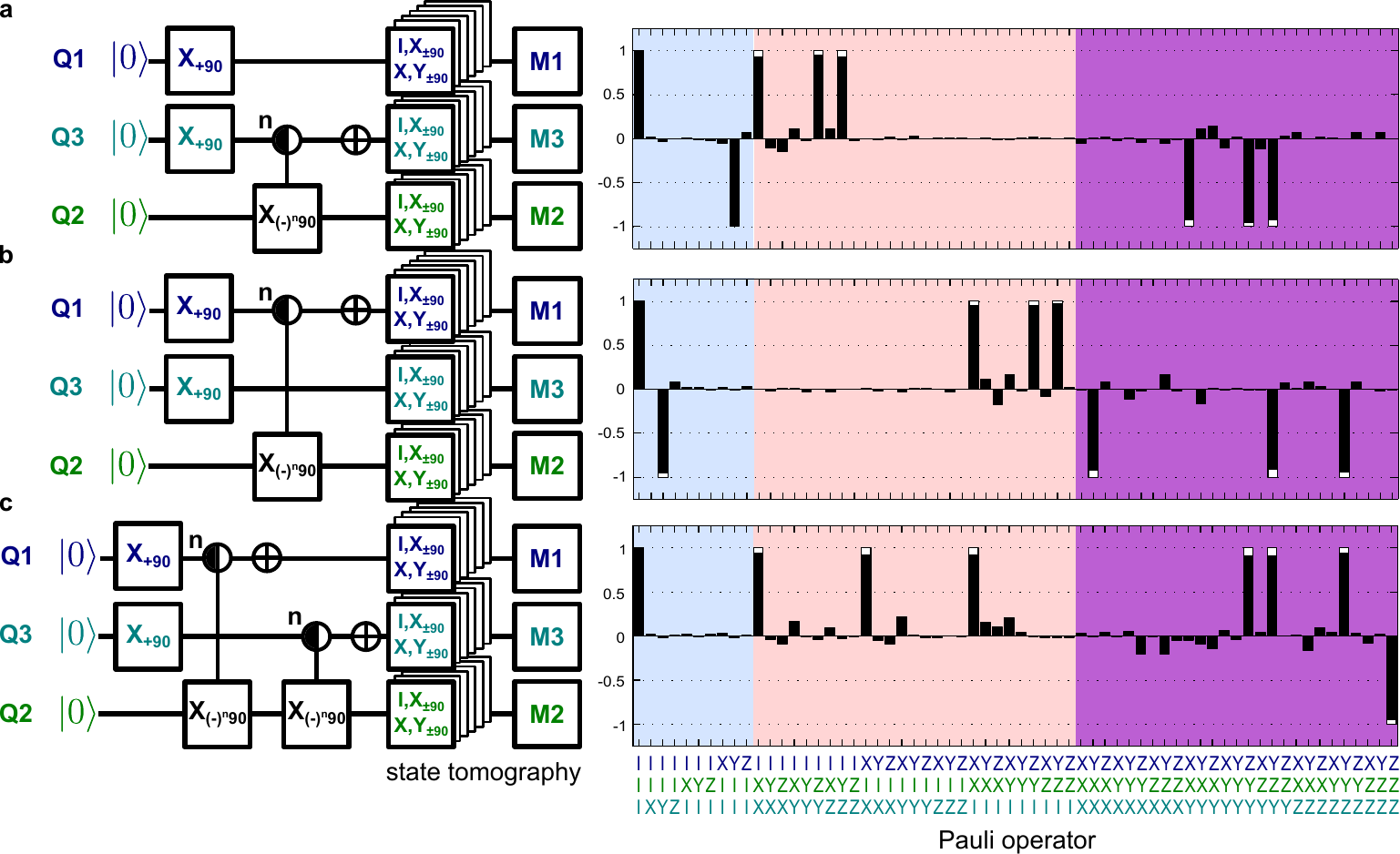}
\caption{\label{fig:3} \textbf{Three-qubit state tomography via correlated single-shot readout.}  Building towards the   action of the PCP on a superposition state of the code qubits (\textbf{c}), we also show the operation of entangling the syndrome Q2 with each of the data qubits Q1 (\textbf{a}) and Q3 (\textbf{b}). Reconstructed three-qubit state represented in Pauli state vector form after entangling syndrome Q2 with, \textbf{a}, code qubit Q3 via a $ZX_{90}$ two-qubit gate (entangled two-qubit state with fidelity $\mathcal{F}_{\text{state}} =0.95 $), \textbf{b}, code qubit Q1 via a $ZX_{90}$ two-qubit gate (entangled two-qubit state with fidelity $\mathcal{F}_{\text{state}} =0.95 $), \textbf{c}, both code qubits Q1 and Q3 via $ZX_{90}$ between Q1 and Q2 and Q3 and Q2 simultaneously (which comprise performing the gate-portion of the PCP) giving the maximally-entangled three-qubit GHZ state ($\mathcal{F}_{\text{state}} = 0.94$). In the shown Pauli vectors, the blue-, pink-, and purple- shaded regions signify single-, two-, and three-qubit Pauli operators, respectively. }
\end{figure*}

Both pairs of $ZX_{90}$ entangling gates are characterized with two-qubit Clifford randomized benchmarking (Extended Data Fig.~\ref{fig:S2}). With a total gate time of 350 ns, the $ZX_{90}$ gate between Q2 and Q1 (Q3) is experimentally shown to have a gate fidelity of $0.962 \pm 0.002$ ($0.957\pm 0.001$). Details about the gate tune-up and calibration can be found in the Methods.

In order to determine the collective state of all qubits in the system, we can perform independent readouts of each qubit through the individually coupled resonators. The syndrome qubit, Q2, is read-out dispersively with the JPA, pumped -4 MHz from the readout frequency ($\omega_{\text{M2}}=2\pi\cdot6.584$~GHz), and an optimized single-shot assignment fidelity (with no preparation corrections) of 91$\%$ is achieved, although fluctuations on the order of 2-3$\%$ are observed. The code qubits are read-out using the high-powered Josephson non-linearity of the readout cavities~\cite{Reed2010}. Further detail and parameters about the readouts are given in the Methods.

\begin{figure*}[htbp!]
\centering
\includegraphics{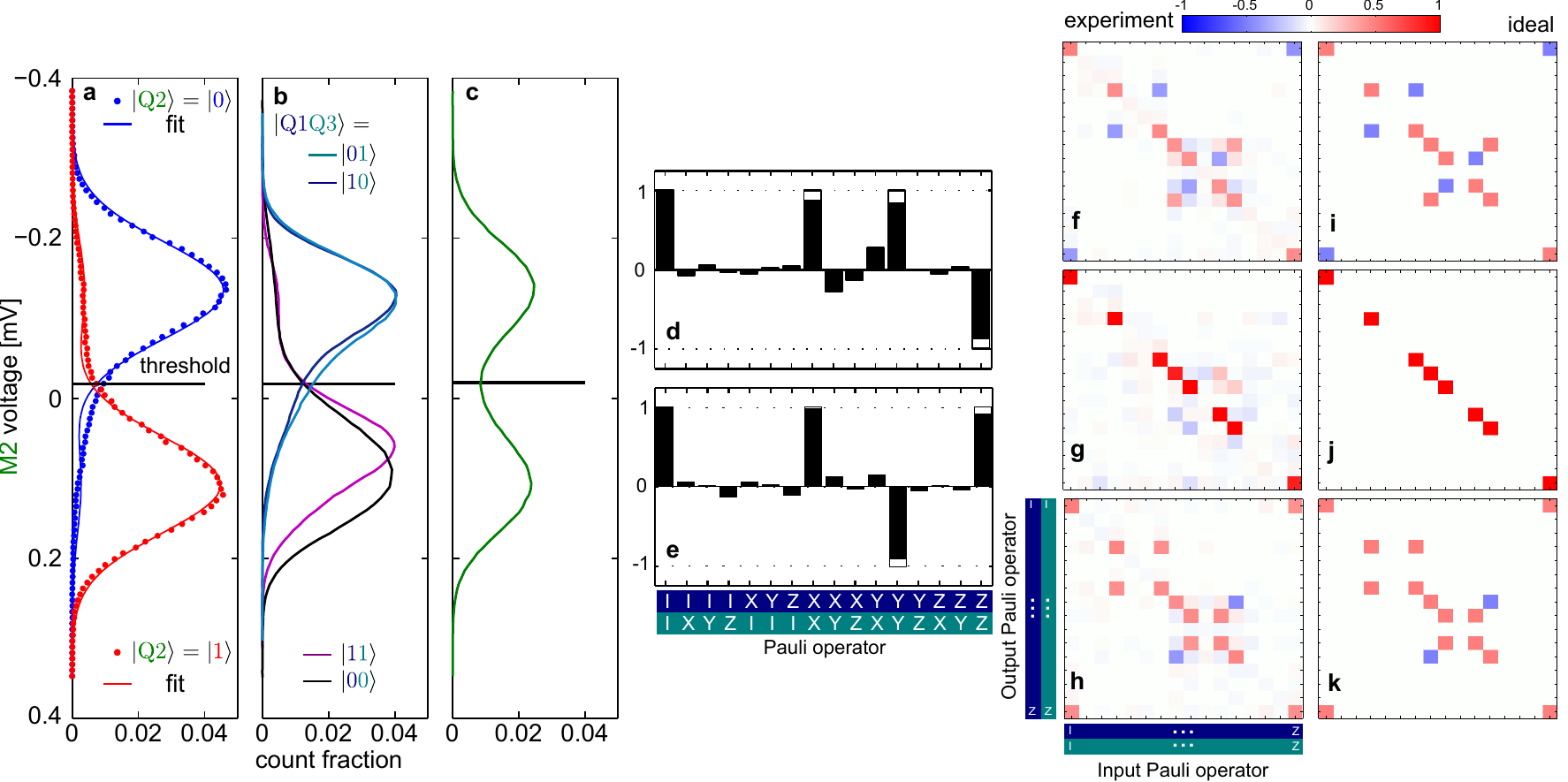}
\caption{\label{fig:4} \textbf{Parity check experiment and non-nearest neighbour pair entanglement generation.} \textbf{a}, Readout histograms of M2 having prepared Q2 in the state $\ket{0}$ and $\ket{1}$ averaged over all basis states for Q1 and Q3. \textbf{b}, Readout histograms of M2 after applying the PCP when Q1 and Q3 are prepared in the four different basis states, $\ket{00}$ (black), $\ket{01}$ (teal), $\ket{10}$ (blue), $\ket{11}$ (purple). \textbf{c}, Readout histogram of measurement M2 after applying the PCP when Q1 and Q3 are prepared in an equal superposition state.   \textbf{d,e}, Pauli state vectors of Q1 and Q3 conditioned on the single-shot measurement of Q2. In the case of Q2 in $\ket{0}$ ($\ket{1}$), state tomography confirms the odd (even) parity Bell state, $\ket{\psi_{\text{odd}}} = (\ket{01} + \ket{10})/\sqrt{2}$  ($\ket{\psi_{\text{even}}} = (\ket{00} + \ket{11})/\sqrt{2}$) with fidelity 0.89 (0.95).  \textbf{f-k}, Pauli transfer matrices of the Z-parity check measurement operation. Top (bottom)  corresponds to the odd (even) projection and has a measurement fidelity of 0.90 (0.91). The central measurement operation corresponds to the unconditional map and has a fidelity of 0.968 with a map that is completely dephasing in the even and odd parity basis.    }
\end{figure*}

To observe the action of the gate portion of the PCP, in which CNOT (in our case $ZX_{90}$) gates are performed between the syndrome and the code qubits, it is insightful to perform tomographic reconstruction of the complete 3-qubit system. State tomography in our system is achieved by correlating individual single shots of all three individual readouts\cite{ryan_inprep2013}, $\text{M}i$, $i\in[1,3]$. Figure~\ref{fig:3} shows reconstructed three-qubit Pauli state vectors for the entanglement processes necessary for the PCP.  In Fig.~\ref{fig:3}\textbf{a} a $ZX_{90}$ entangling operation between the code qubit Q3 and the syndrome Q2 is implemented giving a state fidelity ~$\mathcal{F}_{\text{state}}$ = $0.949\pm 0.002~(\text{sdp}) ~0.954\pm 0.002 ~(\text{raw})$, where sdp refers to a semi-definite program reconstruction of the state and raw reflects unconstrained inversion~\cite{Chow2012PRL}. The difference between the sdp and raw estimates exceeds the statistical error estimated from the readout signal-to-noise and suggests the main source of error in our experiment is systematic.  In Fig.~\ref{fig:3}\textbf{b} a $ZX_{90}$ entangling operation between Q1 and Q2 is implemented ($\mathcal{F}_{\text{state}}$ = $0.951\pm 0.002~(\text{sdp})~ 0.953\pm 0.003~(\text{raw})$) and finally in Fig.~\ref{fig:3}\textbf{c} both $ZX_{90}$ gates are applied simultaneously, generating a maximally-entangled GHZ state of all three qubits ($\mathcal{F}_{\text{state}}$ = $0.935\pm 0.002~(\text{sdp})~0.942\pm 0.003~(\text{raw})$). We thus show the ability to distribute entanglement across the full network, first between nearest-neighbour qubits Q3 and Q2 or Q1 and Q2, and then across the entire system with the GHZ state, spanning both bus resonators. 

With the demonstrated high-fidelity $ZX_{90}$ gate primitives for the PCP, the next step is to observe the single-shot readout of the syndrome Q2 to signal the parity of the code qubits Q1 and Q3. Starting with the simple computational basis states, $\ket{00}$, $\ket{01}$, $\ket{10}$, and $\ket{11}$, as inputs for the code qubits, we observe the proper parity assignment via the PCP, shown via the M2 histograms in Fig.~\ref{fig:4}\textbf{b}. By thresholding the measurement outcomes of M2 based on readout calibration traces (a typical syndrome readout calibration histogram is shown in Fig.~\ref{fig:4}\textbf{a}), we reconstruct the state of Q1 and Q3 conditioned on M2 using standard quantum state tomography techniques (See Methods). In the case of the four computational states, we obtain fidelity of $\mathcal{F}_{\text{sdp}}=0.984, 0.987, 0.989, 0.909$ and $\mathcal{F}_{\text{raw}}=0.975, 0.989, 0.999, 0.905$ respectively.  

A more complete stress test of the PCP is to observe its function on the maximal superposition state of the code qubits Q1 and Q3. The gate protocol now mimics that of the GHZ-state generation from Fig.~\ref{fig:3}\textbf{c}, and over repeated state-preparations and measurements of the syndrome, M2, we obtain a bi-modal histogram, indicating instances of both parities exist (Fig.~\ref{fig:4}\textbf{c}).  We observe the probabilistic entanglement of either the odd or even Bell states $\ket{\psi_{\text{odd}}} = (\ket{01} + \ket{10})/\sqrt{2}$  or $\ket{\psi_{\text{even}}} = (\ket{00} + \ket{11})/\sqrt{2}$ conditioned on M2. For these conditioned entangled states, we find state fidelities of $\mathcal{F}_{\text{odd}} = 0.891 ~(\text{raw}) ~0.891~ (\text{sdp}) $ and $\mathcal{F}_{\text{even}}=0.970 ~(\text{raw}) ~0.948 ~(\text{sdp})$. 

Finally, in order to characterize the complete ideal projective nature of the PCP, we perform measurement tomography. This is accomplished via quantum process tomography of the code qubits, for which further details are given in the Methods. Conditioned on the measurement of the syndrome M2, we obtain the two maps for the odd and even parity projection operators shown in Fig.~\ref{fig:4}\textbf{f} and \ref{fig:4}\textbf{h} (ideal maps are shown in Fig.~\ref{fig:4}\textbf{i} and \ref{fig:4}\textbf{k}). To quantify the performance of the PCP we introduce a measurement fidelity metric that takes into account the full quantum dynamics of the measurement, including projection and back-action (see Methods for more details). We obtain a measurement fidelity of $\mathcal{F}_{\text{meas}}^{\text{odd}}=0.904\pm 0.002$ and  $\mathcal{F}_{\text{meas}}^{\text{even}}=0.912\pm 0.003$. The loss in measurement fidelity corresponds mostly to the $91 \%$ assignment fidelity of M2, best illustrated by the unconditional map (shown in Fig.~\ref{fig:4}\textbf{g}) having a measurement fidelity $\mathcal{F}_{\text{meas}} = 0.968$. 

It is important to note that all the gates used are calibrated and run to achieve these results without any Hamiltonian corrections for either single-qubit or two-qubit errors. An $X$-parity check is a simple extension through the appropriate application of single-qubit Hadamard pulses on the code qubits. 

The experiment described implements a sub-section of the SC fault-tolerant architecture. By combining high-coherence transmon qubits, high-fidelity nearest-neighbour two-qubit gates, and high-fidelity quantum non-demolition single-shot readout, we use a syndrome qubit to determine the parity of its neighbouring qubits. With this device, we demonstrate the versatility of superconducting qubits in the extended quantum bus architecture for application towards a larger fault-tolerant quantum computing device. Looking ahead, direct extensions to 8-qubit full-plaquette and 13-qubit logical qubit demonstrations will be feasible with existing integration techniques. Overcoming integrated circuit engineering hurdles while preserving long coherence times should pave the way for larger surfaces for quantum error correction.
\section*{Methods Summary}

\subsection*{Device fabrication}
The device is fabricated on a 720\,$\mu$m thick silicon substrate. All superconducting coplanar waveguide resonators are defined via optical lithography and subtractive reactive ion etching of a sputtered niobium film (200\,nm thick). The three single-junction transmon qubits are patterned using electron-beam lithography, followed by double-angle deposition of aluminum, with layer thicknesses of 35\,nm and 85\,nm. Liftoff process is used to form the final junction structure.

\subsection*{Device parameters}
The three transmon qubits ($i \in [1,3]$) have transition frequencies \{$\omega_i$\}$/2\pi$ =  \{5.0388, 5.0080, 5.2286\} GHz, with readout resonators at \{$\omega_{\text{R}i}$\}$/2\pi$ = \{6.698, 6.585, 6.695\} GHz, relaxation times \{$T_{1(i)}$\} = \{24, 29, 20\} $\mu$s, \{$T_{2(i)}^{\text{echo}}$\} = \{32, 25, 18\} $\mu$s. The bus resonators are un-measured but $R_{12}$ ($R_{23}$) is designed to resonate at 8 (8.5) GHz. The dispersive cavity shifts of the readout resonators are measured to be \{$\chi_i$\}/$\pi$ = \{-2.0, -2.0, -2.3\} MHz and the readout resonators have line-widths  \{$\kappa_{\text{R}i}$\}$/2\pi$ = \{443, 976, 793\} kHz. All qubits have measured anharmonicities of $-340$~MHz. From the above we calculate coupling strengths 
\{$|g_i|$\}$/2\pi$ = \{70, 67, 67\} MHz to the
readout resonators, which is consistent with electromagnetic simulations.

\section*{Methods}

\subsection*{Experimental setup}

The half-plaquette device is cooled to 15 mK in an Oxford Triton dilution refrigerator. A full schematic of the wiring and experimental control hardware is depicted in Extended Data Fig.~\ref{fig:S1}. Each qubit has its own dedicated readout line with an associated set of isolators and Caltech HEMT (noise temp $\sim$6K) amplifiers. Q2 is unique in that its readout signal is reflected off of a UC Berkeley JPA before going onto the isolator and HEMT chain. The device is housed in a light-tight Ammuneal cryoperm-shield which is coated throughout with a layer of lossy eccosorb (Emerson \& Cuming CR-124). Besides explicit cryogenic attenuators at the different stages of the cryostat, all qubits are also attenuated at the lowest temperature stage with in-house eccosorb coaxial filters. 

Outside the cryostat, all microwave qubit control signals are generated via vector modulation combining off-the-shelf electronics. The microwave readout signals are pulse modulated using Arbitrary Pulse Sequencers built by Raytheon BBN Technologies. The readout signals are processed via two Alazartech ATS9870. All single-shot readout traces are processed with an optimal quadrature rotation filter, described in parallel work~\cite{ryan_inprep2013}.

\begin{suppfigure*}[htbp!]
\centering
\includegraphics{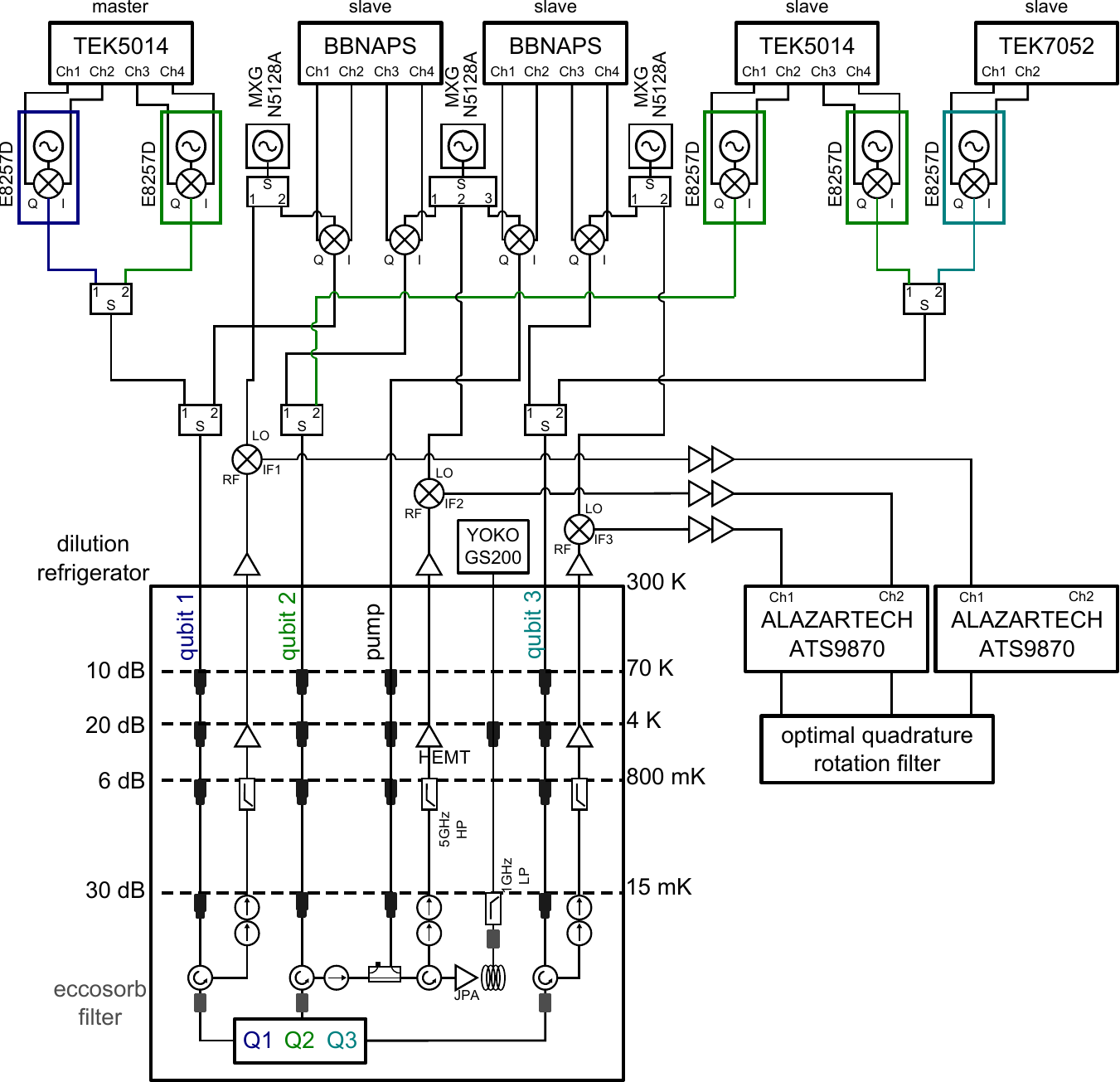}
\caption{\label{fig:S1} \textbf{Detailed schematic of experimental setup.} Wiring scheme for all room temperature controls as well as internal configuration of Oxford Instruments Triton dilution refrigerator.}
\end{suppfigure*}

\subsection*{Calibration sequences}
Complete tune-up of all microwave gates is accomplished using sets of automated repeated sequences. For single-qubit gates, the repeated calibration sequences are described in a previous publication~\cite{Chow2012PRL}.

The cross-resonance pulse amplitude is calibrated in close analogy to single qubit amplitude calibrations. An odd number $2N-1$ of $ZX_{90}$ pulses are applied and the amplitude is adjusted so that for each $N$ the expected signal is halfway between 0 and 1. Any amplitude miscalibrations lead to departures from this expected signal and are amplified for increasing $N$.

In addition to amplitude we must also calibrate the phase of the $ZX_{90}$ pulse between Q3 and Q2 (as well as Q1 and Q2). In our experiment we use a separate microwave generator to supply the cross-resonance pulse on Q3 at the frequency of Q2. The phase of this microwave signal must be calibrated to match that of the microwave generator supplying the single qubit pulses on Q2. This is done by applying the pulse sequence $IY_{90}(ZU_{180}IX)^N IX_{90}$. The $U$ denotes the rotation axis defined by the second generator and the goal is to calibrate for an $X$ rotation. In the case of an $X$-rotation we expect the signal to be halfway between 0 and 1 for each $N$ and miscalibrations of the phase lead to deviations that are amplified with increasing $N$. 
These methods provide a routine for automated calibration with high precision. In the experiments all cross-resonance pulses were calibrated on a regular basis because of phase drift between the two microwave generators. 

\subsection*{Randomized benchmarking}
All single-qubit gates are 40 ns Gaussian-shaped microwave pulses (Gaussian width $\sigma = 10$ ns) resonant with the transition frequencies of the qubits, with scaled derivative-of-Gaussian shapes applied on the quadrature channel to minimize leakage effects~\cite{Motzoi:2009fx}. The gates are all autonomously calibrated with a set of repeated pulse experiments, correcting for: amplitude of $X_{90}$ and $X$ gates, amplitude imbalance between $X$- and $Y$- rotations, mixer skew, and derivative of Gaussian shape parameter. Single-qubit gates are all independently characterized via Clifford~\cite{Magesan2011} randomized benchmarking (RB), and summarized in Table~\ref{table:S1}. To characterize the addressability error of the system, we perform simultaneous~\cite{gambetta_characterization_2012} RB, applying different sets of randomized single-qubit Clifford gates to all three qubits at the same time. These results are also summarized in Table~\ref{table:S1} and essentially indicate that addressability errors are at the 0.1\% error level.

The two-qubit $ZX_{90}$ gates for both pairs of qubits are shaped with Gaussian turn-on (3$\sigma$, $\sigma = 24$ ns), a flat section, and then a Gaussian turn-off, for a total gate time of 350 ns. The $ZX_{90}$ gates are tuned-up also using repeated pulse experiments (described in previous section). It is also important to note that the pair of two-qubit gates can be applied simultaneously, as they commute with one another. To characterize the gates, we generate two-qubit Clifford operations~\cite{corcoles_process_2013} and perform RB. The results for the two cases are shown in Extended Data Fig.~\ref{fig:S2}, where we show the average fidelity decay over 35 different randomized two-qubit Clifford sequences. Analyzing the decay curves gives us error per two-qubit Clifford gate of $0.058 \pm 0.003$ for the Q1 and Q2 and $0.065\pm 0.002$ for Q3 and Q2. We find the reduced chi-square for
these fits are 0.583 and 0.385 respectively. This demonstrates that the model is a faithful representation of the data. As each two-qubit Clifford gate is composed of 1.5 $ZX_{90}$ generators, we estimate the two-qubit $ZX_{90}$ gate errors to be 3.8\% and 4.3\%.

\begin{table}
\begin{ruledtabular}
\begin{tabular}{|c|c c c|}
Randomize & M1 [$\times 10^{-3}$] & M2 [$\times 10^{-3}$] & M3  [$\times 10^{-3}$]\\ 
\hline
Q1 	& $3.06\pm0.05$  	& -- 	& --\\
Q2 	& --		& $2.30\pm0.05$ & --\\ 
Q3	& --		& --	& $2.77\pm 0.05$\\
Q1, Q2, Q3	& $3.8\pm 0.1$ &	$3.32\pm 0.09$ &	$2.89\pm0.05$
\end{tabular}
\end{ruledtabular}
\caption{\label{table:S1} \textbf{Summary of single-qubit randomized benchmarking.}}
\end{table}

\begin{suppfigure}[htbp!]
\centering
\includegraphics{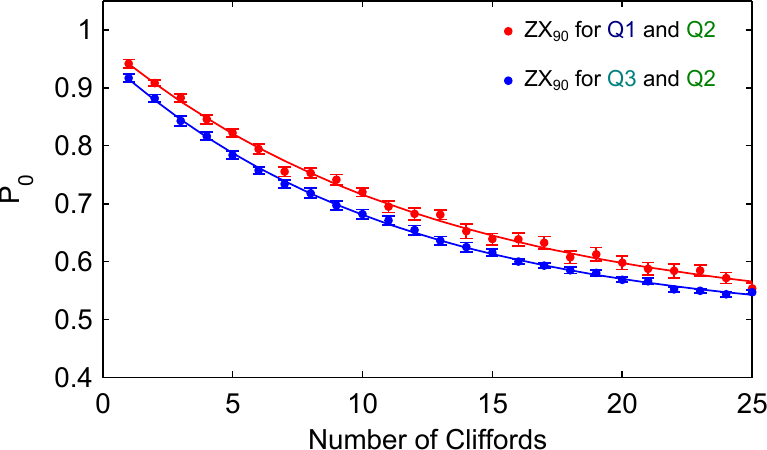}
\caption{\label{fig:S2} \textbf{Two-qubit randomized benchmarking.} Average $P_0$, population of Q2 ground state, versus number of two-qubit Cliffords generated via $ZX_{90}$ gates between Q1 (Q3) and Q2 are shown as red (blue) circles. Experiments are performed randomizing over 35 different sequences of Cliffords. Fits to the RB experiment for Q1 (Q3) and Q2 are shown as solid red (blue) lines, from which we extract an error per two-qubit Clifford of 
 $0.058 \pm 0.003$ ($0.065\pm 0.002$).}
\end{suppfigure}

\subsection*{Readout characterization}

For this experiment each qubit has its own measurement resonator. On Q1 and Q3 high-power readout was used and for Q2 a dispersive linear readout with a JPA was used. The readout was performed by using an integrating kernel that takes into account the response of the cavity (see Ref. \cite{ryan_inprep2013} for more details). This is important when most of the information is in the initial transients of the signal. The integration time for the experiment was 4 $\mu$s for the high power readout and 2 $\mu$s for the dispersive readout with the JPA.

\begin{suppfigure}[htbp!]
\centering
\includegraphics[width=0.45\textwidth]{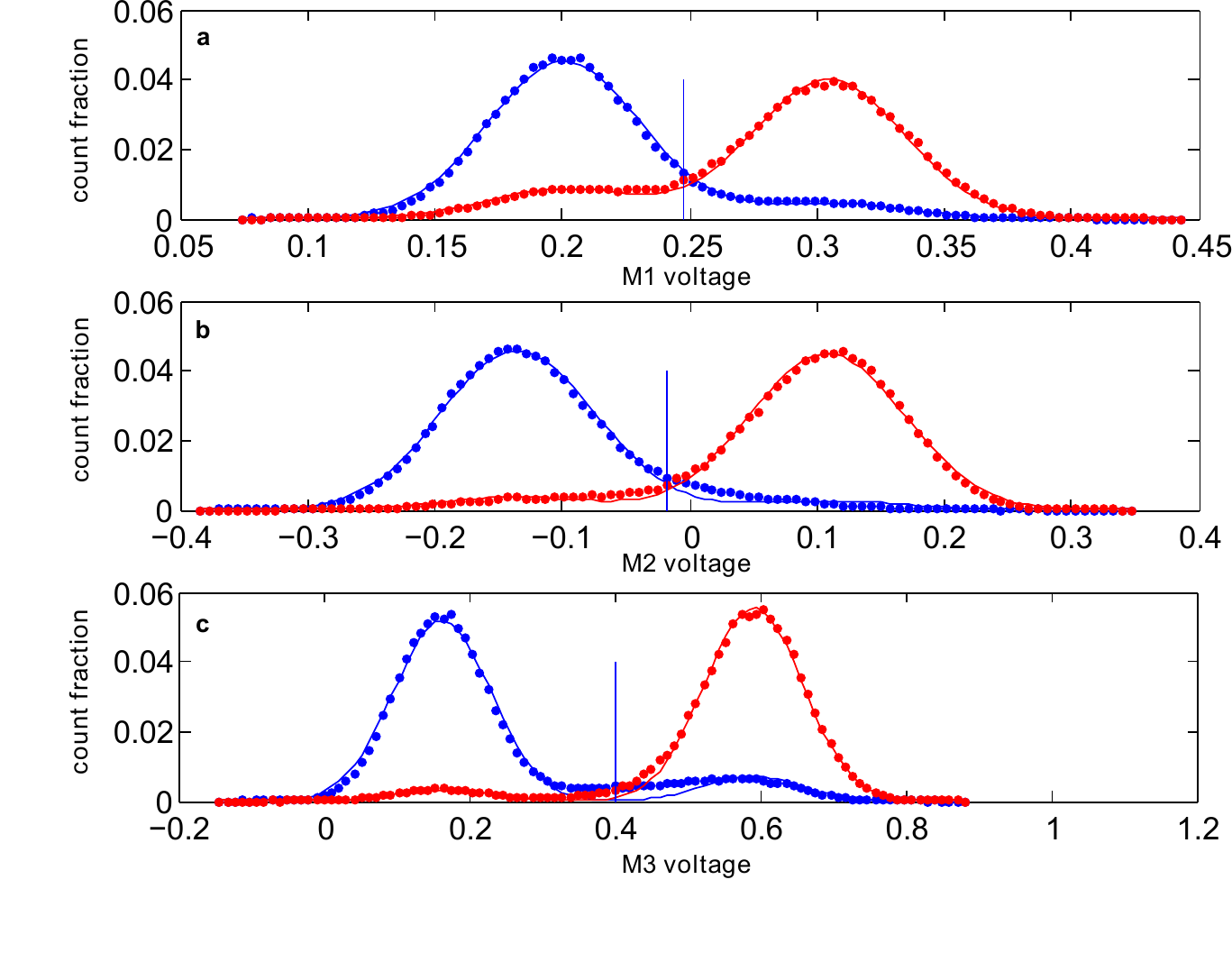}
\caption{\label{fig:S3} \textbf{Histogram of three independent readouts} \textbf{a}, Histograms of the high-power readout used for Q1. \textbf{b}, Histograms of the dispersive linear readout with a JPA  for Q2, \textbf{a}, Histograms of the high-power readout used for Q3. In all cases blue is preparing ground and red is excited. Solid lines are double-gaussian fits to the histograms. }
\end{suppfigure}

Shown in Extended Data Fig. \ref{fig:S3} are typical histograms for the three readout channels averaged over all computational basis for the qubits not measured. Here we see that the assignment fidelity, defined by 
\begin{equation}
\mathcal{F}_{a} = 1 -P(0|1)/2-P(1|0)/2
\end{equation} for the three channels is $0.84$, $0.91$ and $0.89$ respectively. These are typical values and we see about a $2-3 \%$ fluctuation over the course of a typical experiment. By fitting a double Gaussian model to the data we find that the ratio of the undesired state to the desired state for Q1 prepared in the ground (excited) is $9.9\%$($22\%$) for Q2 $5.7\%$($8.3\%$) and for Q3 $13.7\%$($6.0\%$).  We believe most of the error is due to the high-power non-linear readout of Q1 and Q3 and is not due to the qubits being initialized in the wrong state. With no power applied to the Q1 and Q3 resonator the assignment fidelity is $0.95$ and the ratios of the two Gaussians are $5.6\%$ when Q2 is prepared in the excited state and negligible when Q2 is prepared in the ground state.  

\subsection*{State tomography}

For state tomography we used the correlation method as described in Ref. \cite{ryan_inprep2013}. The single shots for each measurement resonator are correlated and from a set of complete post-rotations we can use either linear inversion or a semi-definite program (with constraints $\rho\succeq 0 $ and $\mathrm{tr}(\rho)=1$) to reconstruct the state.  The complete set of rotations used are $\{I, X, X_\mathrm{+90},X_\mathrm{-90},Y_\mathrm{+90},Y_\mathrm{-90}\}^{\otimes n}$. 

Typically 20,000 shots for each post rotation are used and we find that the statistical error in the measured voltages has $\mathrm{SNR}\sim 1\times 10^4,2\times 10^4,2\times 10^4$ for the three measurement channels M1, M2, M3 respectively. The second order correlators range in SNR from $2\sim 5\times 10^3$ and the third order has SNR $\sim 1000$. Using these and a bootstrapping method \cite{Chow2012PRL} we can estimate the state fidelity and the statistical error.
The state fidelity is given by
\begin{equation}
\mathcal{F}_\mathrm{state} = \left(\mathrm{Tr}[\sqrt{\sqrt{\rho_\mathrm{ideal  }}{\rho_\mathrm{noisy  }}\sqrt{\rho_\mathrm{ideal  }}}]\right)^2 , 
\end{equation} where $\rho_\mathrm{ideal  }$ is the ideal state and $\rho_\mathrm{noisy  }$ is the reconstructed state.

 We find that in all cases the fluctuations in the state fidelity from statistics is much smaller than the difference between the linear reconstruction and the semi-definite program. Furthermore, we find typically the sum of all the negative eigenvalues in the three-qubit space to be less than 0.03.

\subsection*{Measurement tomography}

An ideal Z-parity check can be described by the quantum operation
\begin{equation}
\Pi (\rho) = \Pi_\mathrm{even} \rho \Pi_\mathrm{even}\otimes\ket{1}\bra{1} + \Pi_\mathrm{odd} \rho \Pi_\mathrm{odd}\otimes\ket{0}\bra{0},
\end{equation}
where  
\begin{equation}
\Pi_\mathrm{even} = (II + ZZ)/2 ~~~~~~ \Pi_\mathrm{odd} = (II - ZZ)/2
\end{equation} and the extra system is used to label the outcome of measurement of the syndrome qubit. In the noisy case this is represented by the operation 
\begin{equation}\label{eq:con}
\mathcal{E} (\rho) = \mathcal{E}_\mathrm{even} (\rho) \otimes\ket{1}\bra{1} + \mathcal{E}_\mathrm{odd} (\rho)\otimes\ket{0}\bra{0}
\end{equation} and the goal of measurement tomography is to determine the conditional maps $\mathcal{E}_\mathrm{even}(\rho)$ and $\mathcal{E}_\mathrm{odd}(\rho)$. These quantum operations are completely positive but not trace-preserving. 

By binning the results of the measurement on the syndrome qubit, tomography on the two-qubit subspace is performed by preparing a complete set of different input states and measurement bases via pre and post-rotations, and reconstructing the operations from the measurement results. The complete set of rotations that we use are the same as those used in state tomography.  We use both a linear reconstruction and a minimization to make the maps physical. For more details on how quantum process tomography can be performed see Ref.~\cite{Chow2012PRL}.

We use the Pauli transfer matrix \cite{Chow2012PRL} defined by
\begin{equation}
\mathcal{R}_\Lambda(i,j)=\tr(P_i\mathcal{E}(P_j))/d
\end{equation}
to represent the measurement operations where $P_j$ are the standard Pauli operators $\{I,X,Y,Z\}^{\otimes 2}$. 

To quantify the measurement we define the measurement fidelity by a generalization of the average fidelity. Since the measurement maps are not trace preserving we need to use normalized outputs $ {\Pi}'_x(\psi)$, and $ \mathcal{E}'_x(\psi)$, where $\mathcal{A}'_x(\psi)={\mathcal{A}_x(\ket{\psi}\bra{\psi})}/{\|\mathcal{A}_x(\ket{\psi}\bra{\psi})\|_\mathrm{tr}}$, $\mathrm{tr}$ refers to the trace norm, and $x=\{\mathrm{even},~\mathrm{odd}\}$. Doing this gives
\begin{equation}
\mathcal{F}^x_\mathrm{meas}=\int d\psi \tr\left(\sqrt{ \sqrt{{\Pi}'_x(\psi)} \mathcal{E}'_{x}(\psi)\sqrt{{\Pi}'_x(\psi)} 
}\right).
\end{equation} Since the nullspace of a projection operation has measure zero and the noisy realization typically will also have a nullspace of zero measure this integral is well defined. To compute this we draw 150,000 different random states from the 
Fubini-Study measure and compute the average. 

One could also define a process fidelity by computing the state fidelity between normalized Choi matrices of the ideal and noisy operations
\begin{equation}
\mathcal{F}^x_\mathrm{pro} = \frac{\left(\mathrm{Tr}\left[\sqrt{\sqrt{\rho^x_\mathrm{ideal  }}{\rho^x_\mathrm{noisy  }}\sqrt{\rho^x_\mathrm{ideal  }}}\right]\right)^2}{{\mathrm{Tr}[\rho^x_\mathrm{ideal  }]\mathrm{Tr}[\rho^x_\mathrm{noisy  }]}}, 
\end{equation} however for non-unitary processes there is no simple relationship between them. 

The unconditional map can be defined by tracing Eq. (\ref{eq:con}) over the syndrome qubit giving  
\begin{equation}\label{eq:uncon}
\mathcal{E} (\rho) = \mathcal{E}_\mathrm{even} (\rho) + \mathcal{E}_\mathrm{odd} (\rho).
\end{equation} Since this is a quantum operation the standard fidelity between quantum operations can be used.

\section*{Note}

During the completion of this manuscript, we became aware of similar work by O.~P.~Saira et al.~\cite{dicarlo_parity_arxiv}.

\smallskip\noindent
\textbf{Acknowledgments}
	We thank M.~B.~Rothwell and G.~A.~Keefe for fabricating devices. We thank J.~R.~Rozen and J.~Rohrs for experimental contributions, and D.~DiVincenzo, J.~A.~Smolin, and G.~Smith for engaging theoretical discussions. We thank I.~Siddiqi for providing the Josephson Parametric Amplifier. We acknowledge Caltech for HEMT amplifiers. We acknowledge support from IARPA under contract W911NF-10-1-0324. All statements of fact, opinion or conclusions contained herein are those of the authors and should not be construed as representing the official views or policies of the U.S. Government.

\smallskip\noindent
\textbf{Author contributions} J.M.C. and J.M.G. designed the experiments. D.W.A. and J.M.G performed simulations of devices. J.M.C., D.W.A., and S.J.S. characterized devices. J.M.C., J.M.G., A.W.C, and E.M. interpreted and analyzed the experimental data. N.A.M. set-up the microwave control hardware. E.M. and J.M.G. developed the measurement tomography formalism. B.R.J. and C.A.R. developed readout hardware and analysis for correlated readouts. All authors contributed to the composition of the manuscript. 

\smallskip\noindent
\textbf{Author information} The authors declare no competing financial interests. Correspondence and requests for materials should be sent to Jerry M. Chow, chowmj@us.ibm.com.

\clearpage

\end{document}